\documentclass[prd,amsfonts,twocolumn,nofootinbib,showpacs]{revtex4}

\usepackage{graphicx}
\usepackage{dcolumn}
\usepackage{bm}

\def\sigv{\langle\sigma v\rangle}
\def\mchi{m_\chi}

\pacs {26.35.+c, 98.80.Cq, 98.80.Ft}

\begin{document}


\title{Updated CMB constraints on Dark Matter annihilation cross-sections}

\author{Silvia Galli$^{a,b}$, Fabio Iocco$^{c}$, Gianfranco Bertone$^{c,d}$, Alessandro Melchiorri$^a$}
\affiliation{$^a$ Physics Department and INFN, Universita' di Roma ``La Sapienza'', Ple Aldo Moro 2, 00185, Rome, Italy}
\affiliation{$^b$ Laboratoire Astroparticule et Cosmologie (APC), Universit\'e Paris Diderot, 75205 Paris cedex 13, France}
\affiliation{$^c$ Institut d`Astrophysique de Paris, UMR 7095-CNRS Paris, Universit\'e Pierre et Marie Curie, boulevard Arago 98bis, 75014, Paris, France}
\affiliation{$^d$ Institute for Theoretical Physics, Univ. of  Z\"urich, Winterthurerst. 190, 8057 Z\"urich CH}

\begin{abstract}
The injection of secondary particles produced by Dark Matter (DM)
annihilation at redshift $100\lesssim z\lesssim1000$
affects the process of recombination, leaving
an imprint on Cosmic Microwave Background (CMB) anisotropies. Here we provide a new assessment of the constraints set by CMB data on the
mass and self-annihilation cross-section of DM particles. Our new analysis includes the most
recent WMAP (7-year) and ACT data, as well as an improved treatment of the time-dependent 
coupling between the DM annihilation energy with the thermal gas.
We show in particular that  the improved measurement of the polarization signal places already stringent constraints on light DM particles, ruling out `thermal' 
WIMPs with mass $m_\chi \lesssim 10$ GeV.
 
\end{abstract}


\maketitle
\paragraph*{\bf Introduction. }

Precision measurements of the Cosmic Microwave Background (CMB)
anisotropy in temperature and polarization represent powerful tools to constrain new physics processes (see e.g. \cite{galli2010}).
In particular, the remarkable agreement between the
theoretical description of the recombination process, occurring
at $z_r \sim 1000$, and CMB data, severely 
constrains new sources of ionizing
photons, and more in general any deviation from standard
recombination \cite{peeblesmod}, as recently shown by several
groups of authors (see e.g. \cite{galli,naselski,lewisweller}).

In our previous paper \cite{Galli:2009zc} (hereafter GIBM09), we studied the constraints 
that this analysis can set on the properties of dark matter (DM) particles \cite{Bertone:2010zz,Bergstrom:2000pn,Munoz:2003gx,Bertone:2004pz},  under the assumption that
standard recombination is modified by dark matter annihilation {\it only}.

Here, we present an update of GIBM09, and obtain new constraints on the DM particle annihilation 
cross section and mass, based on more recent data
(WMAP 7-year \cite{wmap7komatsu} and ACT 2008  \cite{act} data), and on a new and more 
accurate parametrization of the coupling of the DM-induced energy shower to the thermal gas.

Our results turn out to be competitive
with constraints of diverse astrophysical nature, such as
radio observation of the galaxy, antiprotons, gamma rays from the Galactic center and Galactic 
halo \cite{Pato:2009fn,Cirelli:2009dv,Bergstrom:2008ag,Bertone:2008xr}, but
with respect to them, they have the advantage of not being affected by large 
astrophysical uncertainties.
In fact, our CMB constraints arise from redshifts in the range $100\lesssim z\lesssim 1000$,
i.e. well before the formation of any sizable gravitationally bound structure, and
they therefore do not depend on highly uncertain parameters related
to structure formation, such as halo shape, concentration or minimal mass.

\paragraph*{\bf Annihilating DM and Recombination.}
We briefly recall here the effect of energy injection from DM annihilation 
on the recombination history (see GIBM09 for further details).
High-energy particles injected in the high-redshift thermal gas by DM annihilation
(or decay)  are typically cooled down to the keV scale by high energy processes 
(see details below); 
once the shower has reached this energy scale, the produced secondary particles can
{\it i}): ionize the thermal gas, {\it ii}): induce Ly--$\alpha$ excitation of the hydrogen and {\it iii}): heat
the plasma; the first two  modify the evolution of the free electron fraction $x_e$,
 the third affects the temperature of baryons.
The rate of energy release $\frac{dE}{dt}$ per unit volume
by a relic self-annihilating DM particle is given by
\begin{equation}
\label{enrateselfDM}
\frac{dE}{dt}(z)= \rho^2_c c^2 \Omega^2_{DM} (1+z)^6 p_{ann}, \, \, \,
p_{ann} \equiv f(z) \frac{\sigv}{m_\chi}
\end{equation}
\noindent
where $\sigv$ is the effective self-annihilation rate and $m_\chi$ the mass
of the DM particle, $\Omega_{DM}$ the DM density
parameter and $\rho_c$ the critical density of the Universe today;
the parameter $f(z)$ indicates the fraction of energy which is absorbed
{\it overall} by the gas, under the approximation that the energy absorption
takes place locally. We note that the presence of the brackets in $\sigv$
denote a thermal average, as appropriate for relativistic particles at 
decoupling. At the redshifts of interest here ($z\sim 1000$) the relative velocities of DM 
particles are $v \sim v_{-8} \equiv 10^{-8} c$, i.e. in the extreme non-relativistic limit. Though
holding for s-wave
annihilations, $\sigv_{-8} \simeq \sigv_{dec}$, the same is not true in 
general. For instance,  $\sigv_{-8} \gtrsim \sigv_{dec}$ in models 
with so-called Sommerfeld enhancement, and $\sigv_{-8} \lesssim \sigv_{dec}$ in models with
p-wave annihilations (e.g.  \cite{Bertone:2010zz,Bergstrom:2000pn,Munoz:2003gx,Bertone:2004pz}).

In GIBM09, we considered the fraction of energy $f(z)$ absorbed by the plasma to be constant with redshift, $f(z)=f$. In the following sections, we will present  updated constraints obtained by supposing that $f$ is constant with redshift, as well as constraints considering the actual DM model dependent redshift shape of $f(z)$, as calculated in Slatyer et al \cite{Slatyer:2009yq}.

The formalism we use to introduce the extra energy terms in the recombination equations are the same as in GIBM09, but here we additionally consider the modifications to both helium and hydrogen recombination and we change the extra ionization term  (Eq. 6 in GIBM09). In this equation we do not include Peebles' $C$ factor, since we assume that the extra ionization photons
only ionize the ground state.
 We checked that excluding the factor $C$ in this term in any case does not change the constraints by more than $6\%$ in the case of a constant $f$. The extra Lyman alpha term remains the same as in Eq. 7 of GIBM09.

The most remarkable effect of injecting energy around the recombination epoch
is that the amount of free electrons that survive at low redshift after recombination is larger compared 
to the standard case where no annihilation happens. 
The CMB spectra are therefore affected as follows
(see Ref. \cite{padfink2005} for a more detailed discussion):
the enhanced amount of $x_e$ at low redshift increases the width of last scattering surface, 
and consequently the width of the visibility function. 
This results in a suppression of the amplitude of the oscillation peaks in the temperature and polarization power spectra, especially at scales smaller than the width of the last scattering surface.

 This effect is degenerate with that of other cosmological parameters, affecting the amplitude of peaks at low/high multipoles, such as  the scalar spectral index $n_s$ and the baryon density $\omega_b\equiv\Omega_bh^2$.
In particular, a value of $p_{ann}$ different from zero can be compensated by a higher value of the scalar spectral index $n_s$, that gives more power to smaller scales of the spectrum and, to a much smaller extent, by a higher value of the baryon density $\omega_b$. Nevertheless, this last parameter changes the relative amplitudes of the peaks and is therefore less degenerate with $p_{ann}$ than $n_s$. A smaller degeneracy is also found with the dark matter density $\omega_c\equiv\Omega_ch^2$.

 On the other hand, a larger width of the recombination epoch due to DM annihilation increases the quadrupole moment of the radiation field as well, enhancing the amplitude of the polarization power spectrum at large scales. Furthermore, the fractional contribution to the quadrupole of the temperature monopole with respect to the temperature dipole is increased, therefore slightly shifting the position of the peaks of the polarization power spectrum  (see \cite{padfink2005} for further details). An accurate measurement of the polarization power spectrum can therefore help breaking the degeneracy with $n_s$.

\paragraph*{\bf Updated CMB constraints}

Following GIBM09, 
we compute here the theoretical angular power in presence of DM annihilations,
by modifying the RECFAST routine \cite{recfast}, along the lines described in the
previous section, in the CAMB code \cite{camb}, and by making use of 
package \texttt{cosmomc} \cite{Lewis:2002ah} for Monte-Carlo parameter estimation (see GIBM09 for the details of our statistical analysis, including convergence tests and priors). 
The fractions of energy release inducing ionization, Lyman-$\alpha$ excitation or heating of the baryonic gas are the same as in Chen and Kamionkowski's paper \cite{CK04}, based on the results of Shull and Van Steenberg \cite{ShullVanSteen1985}.  
The dependence on the properties of the DM particles is encoded in the quantity $p_{ann}$,
appearing in eq. \ref{enrateselfDM}, that we use as a parameter in the code.
We start by considering $f$ as a constant parameter, in order to quantify the 
impact of the new CMB data sets on the analysis presented in GIBM09.

Besides $p_{ann}$, we sample the following six-dimensional set of cosmological parameters (with flat priors):
the physical baryon and CDM densities, $\omega_b$ and
$\omega_c$, the scalar
spectral index, $n_{s}$,
the normalization, $\ln10^{10}A_s(k=0.002/Mpc)$,
the optical depth to reionization, $\tau$, and the Hubble parameter $H_0$. 
We consider purely adiabatic initial conditions.

We include the seven-year WMAP data (WMAP7) \cite{wmap7komatsu}(temperature
and polarization), and the 2008 ACT telescope data  \cite{act} (temperature only), that probes the temperature angular power spectrum at small scales ($600\lesssim l\lesssim 8000$, but in our analysis we consider $l_{max}=3500$). The routines for computing the likelihoods for each experiment are supplied by the WMAP and ACT teams respectively. 

\begin{table*}[!t]
\begin{center}
\begin{tabular}{rcccc|cc}
\hline
\hline
& \multicolumn{2}{c}{WMAP7} &\multicolumn{2}{c|}{WMAP7+ACT} & WMAP7 Standard &WMAP7+ACT Standard\\
\hline
\\
$p_{ann} [cm^3/s/GeV]$ & $<2.42\times10^{-27}$& &$<2.09\times10^{-27}$& &-&-\\
$n_s$ &$0.977\pm0.015$& & $0.971\pm0.014$& &$0.963\pm0.014$&$0.962\pm0.013$\\
$100\Omega_bh^2$&$2.266\pm0.057$& &$2.237\pm0.053$& &$2.258^{+0.057}_{-0.056}$&$2.214\pm0.050$\\
$\Omega_ch^2$   &$0.1115\pm0.0054$& &$0.1119\pm0.0053$& &$0.1109\pm0.0056$&$0.1127\pm 0.0054$\\
\hline
\hline
\end{tabular}
\caption{Constraints on the annihilation parameter $p_{ann}$  and on the cosmological parameters that are more degenerate with it, i. e. the scalar spectral index $n_s$, the baryon density $\omega_b$ and the dark matter density $\omega_c$. We report the results using WMAP7 data and WMAP7+ACT data. The constraints on $p_{ann}$ are upper bound at 95\% c.l., while for the other parameters we show the marginalized value and their errors at 68\% c.l. The last two columns reports the value of the cosmological parameters in the standard $\rm \Lambda CDM$ case with no annihilation, as found by the WMAP7 team \cite{wmap7larson} and the ACT team \cite{ACTdunkley}.}
\label{tab:panncost}
\end{center}
\end{table*}

The results of our analysis are in Table \ref{tab:panncost}, there we show the constraints on the DM annihilation parameter $p_{ann}$ and on the parameters that are more degenerate with it (i.e. $n_s,\omega_b,\omega_c$) obtained using WMAP7 data and the combination of WMAP7 plus ACT data. 
We also report the  the constraints obtained by the WMAP7 team \cite{wmap7larson} and by the ACT team \cite{ACTdunkley} on the cosmological parameters in the standard case (no DM annihilation) in order to show the bias introduced  on the cosmological parameters by not considering DM annihilation in the analysis. 

Note in particular that the constraint on $p_{ann}$ with WMAP7 data is improved by a factor $\sim1.8$
with respect to the  WMAP5 constraint obtained in GIBM09.
The 7-year data release has in fact a better measurement of the third peak of the temperature power spectrum at $l\sim 1000-1200$ and of the second dip in the temperature-polarization power spectrum at $l\sim 450$. This allows a better measurement of $\omega_b$ and $\omega_c$ and a partial break of the degeneracy with $p_{ann}$. 
On the other hand, the bias on $n_s$ remains noticeable at $1-\sigma$ level, as the measurement of the polarization power spectrum is still not sufficient to break the degeneracy with $p_{ann}$.
Adding the information at small scales from the ACT data additionally improves the constraint on $p_{ann}$ by $\sim 13\%$. 
As we can see from the table, the Harrison-Zel'dovic model $n=1$ is consistent with the WMAP+ACT analysis in between two standard deviations when dark matter annihilation is considered.

We have also checked our results when other non-standard parameters are considered in addition to the standard ones and $p_{ann}$. We have considered one of the following additional parameters at the time: the fraction of Helium abundance $Y_{He}$, the massive neutrino density, $\Omega_\nu$, and the running of the scalar spectral index $\alpha$. None of these parameters appeared to be degenerate with $p_{ann}$, therefore not affecting the results on the upper limits reported in Table \ref{tab:panncost}.

\begin{figure}[h!]
\begin{center}
\includegraphics[width=0.5\textwidth]{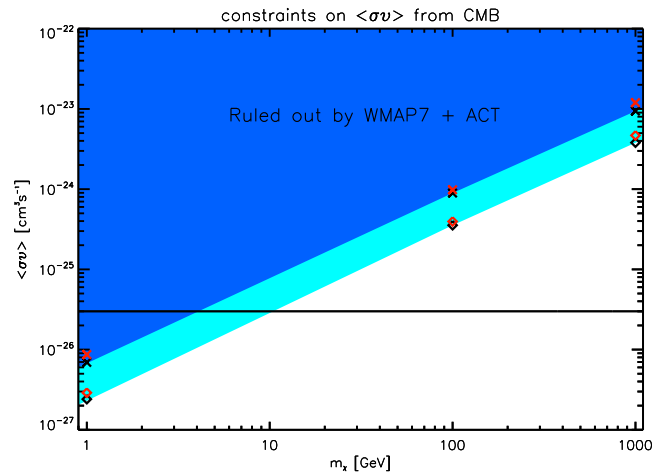}
\caption{Constraints on the cross section $<\sigma v>$ in function of the mass, obtained using a variable $f(z)$ for particles annihilating in muons ($x$ signs) and in electrons (diamonds) using WMAP7 data (red) and WMAP7+ACT data (black) at 95\% c.l.. The exclusion shaded areas 
 are obtained for interpolation of the WMAP7 + ACT data points for muons (dark shading) and electrons (light shading). The black solid line indicates the standard thermal cross-section $<\sigma v>=3\times 10^{-26}cm^3/s$.}
\label{plot}
\end{center}
\end{figure}

\paragraph*{\bf Implementing the redshift dependence of $f$.}
We have so far worked under the assumption that the fraction of the rest DM 
mass energy absorbed by the plasma is constant with
redshift. 
Yet, the fraction of the initial energy deposited into the gas is not constant with cosmic time,
even if the on--the--spot approximation holds true at all redshifts of interest.
This problem has been addressed in \cite{Slatyer:2009yq}, where the authors have computed 
the evolution of the energy fraction $f(z)$ for different primary species, and DM particle mass.
As it can be seen from their Figure 4, the $f(z)$ is a smoothly varying function of redshift 
(even more so for the values of interest in our problem $100\lesssim z\lesssim1000$).
We show the constraints for time-varying $f(z)$ in Figure \ref{plot}. Interestingly, the new results rule out `thermal' WIMPs with mass $m_\chi \lesssim 10$ GeV.

We have checked the constraints which 
is possible to place 
using the redshift dependent shape of $f$ presented in  Equation A1 and Table 1 of 
\cite{Slatyer:2009yq}. 
We have obtained constraints for purely DM  models annihilating
solely (and separately) into electrons and muons, with different DM masses, reported in Table \ref{tab:fvarconstr}. This choice of annihilation channels brackets the possible values of $f(z)$: the case of annihilation to other channels (except of course neutrinos, which practically do not couple at all with the plasma) falls between the two limiting cases studied here.

\begin{table*}[t!]
\begin{center}
 \begin{tabular}{lccc|ccc}
\hline
\hline
&&\multicolumn{2}{c|}{$<\sigma v>$ in $\rm[cm^3/s]$ with Variable $f$}&&\multicolumn{2}{c}{$<\sigma v>$ in $\rm[cm^3/s]$ with Constant $f=f(z=600)$}\\
$\mchi\rm [GeV]$&channel &WMAP7 &WMAP7+ACT&$f(z=600)$&WMAP7&WMAP7+ACT\\
\hline
\hline\\

1 GeV &e$^+$e$^-$&$<2.90\times10^{-27}$&$<2.41\times10^{-27}$&0.87&$<2.78\times 10^{-27}$&$<2.41\times 10^{-27}$\\

100 GeV&e$^+$e$^-$&$<3.95\times10^{-25}$&$<3.55\times10^{-25}$&0.63&$<3.87\times 10^{-25}$&$<3.35\times 10^{-25}$\\
\medskip
1TeV&e$^+$e$^-$ &$<4.68\times10^{-24}$&$<3.80\times10^{-24}$&0.60&$<4.02\times 10^{-24}$&$<3.48\times 10^{-24}$\\

1 GeV &$\mu^+\mu^-$&$<8.68\times10^{-27}$&$<6.93\times10^{-27}$&0.30&$<8.03\times 10^{-27}$&$<6.95\times 10^{-27}$\\

100 GeV&$\mu^+\mu^-$&$<9.82\times10^{-25}$&$<8.94\times10^{-25}$&0.23&$<1.03\times 10^{-24}$&$<8.91\times 10^{-25}$\\

1TeV&$\mu^+\mu^-$&$<1.20\times10^{-23}$&$<9.41\times10^{-24}$&0.21&$<1.15\times 10^{-23}$&$<9.96\times 10^{-24}$\\

\hline
\hline
\end{tabular}
\caption{Upper limits on self-annihilation cross section at   $95 \%$ c.l. using WMAP7 data and a combination of WMAP7 and ACT data . On the left-side of the table, we show the results obtained using the proper variable $f(z)$ for each model. On the right side, for sake of comparison, we show the results obtained by taking the constraints for a constant generic $f$ reported in Table \ref{tab:panncost}, and then calculating $<\sigma v>$ for each case imposing that $f$ is equal to the corresponding $f(z=600)$ for each model. We show results for particles annihilating in electrons and muons.}
\label{tab:fvarconstr}
\end{center}
\end{table*}

Although the implementation of the z-dependence of $f$ clearly leads to more accurate results, we found that 
taking a simplified analysis with constant $f$, such that $f(z=600)=f_{const}$, leads to a difference with respect to the full $f(z)$ approach
of less than $\sim15\%$, depending on the annihilation channel considered. 

\paragraph*{\bf Discussion and Conclusions.}
In this \textit{brief report} we have provided new updated CMB constraints on WIMP annihilations, 
with an improved analysis that includes more recent CMB data (WMAP7 and the ACT2008) and implementing the redshift evolution of the thermal gas opacity to the high energy primary shower. We have also found that a simplified analysis with constant $f=f(z=600)$ leads to an error on the maximum DM self-annihilation cross section smaller than $\sim15\%$, with respect to a treatment that fully takes into account the redshift dependence of $f(z)$.
 
While we were finalizing this paper, Hutsi et al. (HCHR2011) \cite{hutsi} have reported results from a similar analysis, using an averaged evolution of the $f(z)$. 
 They provide  $2-\sigma$ upper limits from WMAP7 with $1-\sigma$ uncertainties on these limits due to the method used. These results are a factor 
between 1.2 and 2 weaker than ours.

This is partially due to the fact that we account for extra Lyman radiation in our code, but this can account for only less than $10\%$ of the difference between the results. 

As in GIBM09, we have calculated how much the Planck satellite and a hypothetical Cosmic Variance Limited experiment will improve the constraints compared to  WMAP7 in the case of constant $f$  (constraints for Planck and CVL reported in GIBM09). We obtain improvement factors of 8 and 23 for Planck and CVL respectively, which are compatible with the ones reported in HCHR2011, 6 and 13. The difference for the CVL experiment is attributed to the slightly different specifications used for the CVL experiment in HCHR2011 and in GIBM09, namely the maximum multipole considered in the analysis, as also stated in HCHR2011.
Clearly the data from the on-going Planck satellite mission, expected to be released by early $2013$, will play a crucial role in constraining additional sources of ionization, such as DM annihilation, in the early universe.

\paragraph*{Acknowledgments.} 
This work is supported by PRIN-INAF, "Astronomy probes fundamental physics".
Support was given by the Italian Space Agency through the ASI contracts "Euclid- IC" (I/031/10/0).
FI is supported from the European Community research program FP7/2007/2013 
within the framework of convention \#235878.

\newcommand\AAP[3]{AAP{\bf ~#1}, #2~ (#3)}
\newcommand\AL[3]{A. Lett.{\bf ~#1}, #2~ (#3)}
\newcommand\AP[3]{Astropart. Phys.{\bf ~#1}, #2~ (#3)}
\newcommand\AJ[3]{Astron. J.{\bf ~#1}, #2~(#3)}
\newcommand\APJ[3]{Astrophys. J.{\bf ~#1}, #2~ (#3)}
\newcommand\ApJ[3]{Astrophys. J.{\bf ~#1}, #2~ (#3)}
\newcommand\APJL[3]{Astrophys. J. Lett. {\bf ~#1}, L#2~(#3)}
\newcommand\APJS[3]{Astrophys. J. Suppl. Ser.{\bf ~#1}, #2~(#3)}
\newcommand\MNRAS[3]{MNRAS{\bf ~#1}, #2~(#3)}
\newcommand\MNRASL[3]{MNRAS Lett.{\bf ~#1}, L#2~(#3)}
\newcommand\NPB[3]{Nucl. Phys. B{\bf ~#1}, #2~(#3)}
\newcommand\PLB[3]{Phys. Lett. B{\bf ~#1}, #2~(#3)}
\newcommand\PRL[3]{Phys. Rev. Lett.{\bf ~#1}, #2~(#3)}
\newcommand\PR[3]{Phys. Rep.{\bf ~#1}, #2~(#3)}
\newcommand\PRD[3]{Phys. Rev. D{\bf ~#1}, #2~(#3)}
\newcommand\SJNP[3]{Sov. J. Nucl. Phys.{\bf ~#1}, #2~(#3)}
\newcommand\ZPC[3]{Z. Phys. C{\bf ~#1}, #2~(#3)}
\newcommand\SCI[3]{Sci.{\bf ~#1}, #2~(#3)}

\end{document}